\documentclass[10pt]{article}
\usepackage{graphicx,amssymb,amsmath}
\newcommand{\be}{\begin{equation}}
\newcommand{\ee}{\end{equation}}
\newcommand{\beq}{\begin{eqnarray}}
\newcommand{\eeq}{\end{eqnarray}}
\def\n{\mathfrak{n}}
\def\s{\mathfrak{s}}
\def\Tr{\mathrm{Tr}}

\def\H{\mathbb{H}}
\def\C{\mathbb{C}}
\def\E{\mathbb{E}}
\newcommand{\mbold}[1]{\mbox{\boldmath{\ensuremath{#1}}}}
\newcommand{\mf}[1]{\mathfrak{#1}}

\begin{document}

\title{Ricci Nilsoliton Black Holes}

\author{ Sigbj{\o}rn  Hervik\\
{\it  \small Dalhousie University, Dept. of Mathematics and Statistics,}\\
{\it  \small Halifax, NS, Canada B3H 3J5}\\ \\
{\small Current address: \it Dept. of Mathematics and Natural Sciences} \\
{\it \small University of Stavanger, N-4036 Stavanger, Norway}\\
{\small E-mail:} {\tt\small sigbjorn.hervik@uis.no}}
\date{\small\today}
\maketitle
\begin{abstract}
We follow a constructive approach and find higher-dimensional black holes with Ricci nilsoliton horizons. The spacetimes are solutions to Einstein's equation with a negative cosmological constant and generalises therefore, anti-de Sitter black hole spacetimes. The approach combines a work by Lauret -- which relate so-called Ricci nilsolitons and Einstein solvmanifolds -- and an earlier work by the author. The resulting black hole spacetimes are asymptotically Einstein solvmanifolds and thus, are examples of solutions which are not asymptotically Anti-de Sitter. We show that any nilpotent group in dimension $n\leq 6$ has a corresponding Ricci nilsoliton black hole solution in dimension $(n+2)$. Furthermore, we show that in dimensions $(n+2)>8$, there exists an infinite number of locally distinct Ricci nilsoliton black hole metrics.  
\end{abstract}

\section{Introduction}
The last decade has seen the interest for negatively curved spaces growing considerably. From a mathematical point of view, the negatively curved spaces have an extremely rich structure; for example, in three dimensions "most" manifolds are negatively curved \cite{thurston,thur:97,BP}. From a physical point of view, negatively curved spaces have arisen both in superstring theories and in higher-dimensional theories of our universe (see e.g., \cite{Mald,Ortin,randall}). The maximally symmetric Anti-de Sitter space (AdS)-- which is a solution to the Einstein equations with a negative cosmological constant -- is the space that has attracted the most attention. In this paper, however, we will draw attention to some other negatively curved solutions to Einstein equations with a negative cosmological constant.  In the mathematics literature they are known as Einstein solvmanifolds \cite{Wolter,Wolter2,Heber}, and unlike the AdS spaces, are not maximally symmetric. We will study a class of Einstein solvmanifolds and show that they allow for a simple generalisation which can be interpreted as black hole solutions with a horizon geometry being that of a nilmanifold. 

For a Lie algebra, $\mf{g}$,  we can contruct the two  descending  series, 
\beq 
\mf{g}^{(0)}_D=\mf{g},&& \quad \mf{g}^{(i+1)}_D=[\mf{g},\mf{g}^{(i)}_D],\nonumber \\ 
\mf{g}^{(0)}_C=\mf{g},&& \quad \mf{g}^{(i+1)}_C=[\mf{g}^{(i)}_C,\mf{g}^{(i)}_C], \nonumber
\eeq
called the derived and the lower central series, respectively. If the derived series terminates, i.e., $\mf{g}^{(k)}_D=0$ for an integer $k$,  we call the Lie algebra $\mf{g}$ \emph{nilpotent}. Similarly, if $\mf{g}^{(k)}_C=0$ for an integer $k$, we call the Lie algebra $\mf{g}$ \emph{solvable}. Clearly, any nilpotent Lie algebra is also solvable. Here, we will denote a generic nilpotent Lie algebra \footnote{The Abelian algebras are trivially nilpotent; however, we will assume that $\n$ is non-Abelian.}  $\n$. Any Lie algebra, $\mf{g}$, gives rise to a unique connected and simply connected Lie group, $G$, such that the tangent space of $G$ (as a manifold) at the unit element is $\mf{g}$: $\mf{g}=T_eG$  \cite{Chevalley}. Any such Lie group can be equipped with a left-invariant metric which turns $G$ into a Riemannian space having a metric which is invariant under the left action of $G$ (see e.g., \cite{Milnor:76}). In the case of a nilpotent Lie algebra $\n$, this gives rise to a nilpotent Lie group $N$. Such a nilpotent Lie group equipped with a left invariant metric is commonly denoted as a \emph{nilmanifold}. We will assume that this metric is Riemannian, unless stated otherwise. 

A Lie group usually possesses \emph{many} non-isometric left-invariant metrics. A natural question would therefore be: Is there a particularly nice or distinguished left-invariant metric? Such a distinguished metric can, for example, be an \emph{Einstein metric} \cite{Besse}; that is, a metric $g_{\mu\nu}$ that obeys
\beq
R_{\mu\nu}=\lambda g_{\mu\nu},
\eeq
where $R_{\mu\nu}$ is the Ricci tensor.
However, this is not appropriate for nilmanifolds since a well known result states that nilmanifolds do not allow for a left-invariant metric which is Einstein \cite{Milnor:76}.
On the other hand, Lauret \cite{L1} noted that some nilpotent groups allow for metrics which obey 
\beq
R_{\mu\nu}=\lambda g_{\mu\nu}+D_{\mu\nu},
\label{eq:nilsoliton}\eeq
where $D^{\mu}_{~\nu}$ as a linear map, ${\sf D}:\mf{n}\mapsto\mf{n}$, is a \emph{derivation of $\mf{n}$}; i.e. 
\[ {\sf D}\left([{ X},{ Y}]\right)=[{\sf D}({X}),{ Y}]+[{ X},{\sf D}({ Y})]. \]
These metrics have a nice interpretation in terms of special solutions of the \emph{Ricci flow} \cite{Hamilton}. For a curve ${\bf g}(t)$ of Riemannian metrics on a manifold $M$, the Ricci flow is defined by the equation 
\beq
\frac{\partial g_{\mu\nu}}{\partial t}=-2R_{\mu\nu}.
\label{eq:Ricciflow}\eeq
If a solution to the Ricci flow (\ref{eq:Ricciflow}) moves by a diffeomorphism and is also scaled by a factor at the same time, we call the solution a \emph{homothetic Ricci soliton} \cite{RicciFlow}. In other words, if $\phi_t$ is a one-parameter family of diffeomorphims generated by some vector field and
\[ {\bf g}(t)=c(t)\phi_t^*{\bf g} \]
is a solution of the Ricci flow, then ${\bf g}$ is  a homothetic Ricci soliton. 

\emph{Ricci nilsolitons} are nilmanifolds with left-invariant metrics being homothetic Ricci solitons. 
In addition, a Ricci nilsoliton has a unique decomposition as given by eq.(2); hence, Ricci nilsolitons are in some way a generalisation of Einstein metrics to nilpotent groups. These Ricci nilsolitons are also unique 
up to isometry and scaling and can therefore be taken to be distinguished left-invariant metrics 
on nilmanifolds \cite{L1}. 
\footnote{Interestingly, the Ricci flow has shown to be of importance in resolving the Thurston geometrisation conjecture \cite{thurston} and thereby the famous Poincar\'e conjecture (see e.g., \cite{Morgan}).}

In this paper we will study black hole solutions where the horizon is locally a nilmanifold, while the total spacetime is a solution to the Einstein equations with a negative cosmological constant. Given an $n$-dimensional nilmanifold $N$, we will see that we can construct such black hole solutions (with compact horizons) of dimension $(n+2)$ provided that 
\begin{enumerate}
\item{} The nilmanifold, $N$, allows for a nilsoliton metric. 
\item{} The nilmanifold, $N$, allows for a compact quotient; i.e., there exists a lattice $\Gamma\subset N$ such that $N\slash\Gamma$ is compact. 
\end{enumerate}
The nilsoliton metric will correspond to the (local) horizon geometry and consequently these solutions are Ricci nilsoliton black holes. 
In particular, we will see that for any nilmanifold of dimension $\leq 6$, both requirements are fulfilled (thus, there exists a corresponding black hole solution of dimension $\leq 8$). On the other hand, for nilmanifolds of dimension $>6$, these requirements are not always fulfilled, however, we will show that there exists an infinite family of nilmanifolds for which they do. This implies that, for spacetime dimension $>8$, there exists an infinite number of locally distinct Ricci nilsoliton black holes. 

The paper is organised as follows. First, we show how to construct Ricci nilsolition metrics through a variational procedure. Then, using a method of Lauret, we construct Einstein solvmanifolds which constitute the "background" spacetime. A simple generalisation allows us to construct black hole solutions having Ricci nilsolitons as horizon geometries. Some aspects of these solutions are discussed, among them, the asymptotic geometry. The paper is constructive in nature and therefore, in the Appendix, a full list of the nilpotent Lie algebras of dimension $\leq 6$ is given, along with their corresponding Ricci nilsoliton solution.

\section{Finding Ricci nilsoliton metrics through a variational procedure} 
Let us consider a vector space $\n$ with a fixed inner product $\langle\cdot,\cdot\rangle$ on $\n$. Define a nilpotent Lie algebra $\mu$ on $\n$ by the structure constants; i.e., 
\beq
[{\sf e}_i,{\sf e}_j]=\mu({\sf e}_i,{\sf e}_j)=\mu^k_{ij}{\sf e}_k, \quad \langle{\sf e}_i,{\sf e}_j\rangle=\delta_{ij}.
\eeq
The set of nilpotent Lie algebras can be considered as an algebraic subset of $V=\wedge^2\n^*\otimes\n$, the vectorspace of all skew-symmetric maps from $\n\times \n$ into $\n$. 
Any nilpotent Lie algebra $\mu$ defines a corresponding simply connected nilpotent Lie group, $N_\mu$, endowed with the left-invariant Riemannian metric determined by $\langle\cdot,\cdot\rangle$. 

Moreover, define the action of ${\bf A}\in GL(n)$ on $\mu$ by 
\beq
{\bf A}*\mu(X,Y)={\bf A}\mu({\bf A}^{-1}X,{\bf A}^{-1}Y),\quad X,Y\in\n.
\eeq
If $\tilde\mu$ and $\mu$ are two Lie algebras, then $\tilde\mu$ and $\mu$ are isomorphic as Lie algebras if and only if they are in the same $GL(n)$ orbit. Furthermore, the corresponding nilmanifolds $N_{\tilde\mu}$ and $N_{\mu}$ are  isometric if and only if they are in the same $O(n)$-orbit.  

The Ricci operator of $N_{\mu}$ can be calculated to be 
\beq
\langle{\bf R}_\mu{\sf e}_i,{\sf e}_j\rangle=\frac 14\sum_{kl}\big[\langle\mu({\sf e}_k,{\sf e}_l),{\sf e}_i\rangle\langle\mu({\sf e}_k,{\sf e}_l),{\sf e}_j\rangle-2\langle\mu({\sf e}_i,{\sf e}_k),{\sf e}_l\rangle\langle\mu({\sf e}_j,{\sf e}_k),{\sf e}_l\rangle\big].
\label{Ricciop}\eeq
Also, consider the two functionals, $R(\mu)$ and $F(\mu)$ defined by 
\beq
R(\mu)&\equiv & \mathrm{Tr}({\bf R}_\mu), \nonumber \\
F(\mu)&\equiv & \mathrm{Tr}({\bf R}^2_\mu).
\eeq
We note that these functionals are the Ricci scalar and the "square" of the Ricci tensor, $R_{ij}R^{ij}$, respectively, of the left-invariant metric. 
The inner product $\langle\cdot,\cdot\rangle$ defines an inner product on $V$, also denoted $\langle\cdot,\cdot\rangle$, by 
\beq
\langle\mu,\tilde\mu\rangle=\sum_{ijk}\langle\mu({\sf e}_i,{\sf e}_j),{\sf e}_k\rangle\langle\tilde\mu({\sf e}_i,{\sf e}_j),{\sf e}_k\rangle=\sum_{ijk}(\mu^k_{ij})(\tilde\mu^k_{ij}). 
\eeq
This inner product defines a natural normalisation of $V$: 
\beq
{\sf S}=\left\{ \mu\in V~\big|~\langle\mu,\mu\rangle=1 \right\}.
\eeq
Note that this normalises $R(\mu)$ since $R(\mu)=-\langle\mu,\mu\rangle/4$. Since a constant rescaling of the metric will rescale $R(\mu)$, there is no loss of generality to restrict to ${\sf S}$. 
 
 It is desirable to find a distinguished metric on a nilmanifold. Since nilpotent groups do not allow for Einstein metrics, we can try the next best thing, namely minimize the functional
\beq
\mathrm{Tr}\left[{\bf R}_\mu-\frac{1}{n}\mathrm{Tr}({\bf R}_{\mu}){\bf 1}\right]^2=F(\mu)-\frac 1n R(\mu)^2,
\eeq
which measures how far $N_{\mu}$ is from being an Einstein space. Therefore, fixed points of $F$ restricted to ${\sf S}$ are of particular significance.
In fact, we have the following theorem by Lauret \cite{L2}:
\paragraph{Theorem:}{\sl  For a nilpotent $\mu\in {\sf S}$ the following statements are equivalent:
\begin{enumerate}
\item{} $N_{\mu}$ is a Ricci nilsoliton.
\item{} $\mu$ is a critical point of $F:{\sf S}\mapsto \mathbb{R}$.
\item{} $\mu$ is a critical point of $F:GL(n)*\mu\cap{\sf S}\mapsto \mathbb{R}$.
\item{} ${\bf R}_{\mu}\in\mathbb{R}{\bf 1}\oplus\mathrm{Der}(\mu)$.
\end{enumerate} }
This Theorem intimately connects the critical points of $F$ and the Ricci nilsolitons. The Ricci nilsoliton metrics can therefore be considered to be  particularly nice metrics on nilpotent groups. 

For a Ricci nilsoliton there exists a symmetric derivation ${\bf D}\in\textrm{Der}(\mu)$ such that 
\beq
{\bf R}_{\mu}=c_{\mu}{\bf 1}+\Tr({\bf D}){\bf D}, \quad \Tr({\bf R}_{\mu}{\bf D})=0.
\label{NilsolitonRicci}\eeq
Thus, 
\[ c_{\mu}=\frac{F(\mu)}{R(\mu)}=-\Tr({\bf D}^2).\]
Since nilmanifolds are never Einstein, we necessarily have ${\bf D}\neq 0$ for a Ricci nilsoliton. Moreover, the scalar curvature never vanishes so $R(\mu)<0$. 

A necessary condition for a Ricci nilsoliton metric to exist for a given nilpotent Lie algebra $\mu$, is therefore that  $\mu$ has a non-zero symmetric derivation ${\bf D}$. In particular, if $\mu$ is \emph{characteristically nilpotent} (i.e., $\mathrm{Der(\mu)}$ is nilpotent) then there cannot exist such a ${\bf D}$ and no Ricci nilsoliton metric exists. Therefore, not all nilmanifolds allow for a Ricci nilsoliton metric. On the other hand, it has been proven that all nilmanifolds of dimension $\leq 6$ allow for one \cite{Will}. 

We now have an algorithm for finding Ricci nilsolitons for a given nilpotent Lie algebra $\mu$ (if it exists): 
\begin{enumerate}
\item{} Find the critical points of $F:GL(n)*\mu\cap{\sf S}\mapsto \mathbb{R}$. Any $\mu\in V$ such that $\mu/\langle\mu,\mu\rangle^{1/2}$ is a critical point will then correspond to a Ricci nilsoliton metric.
\item{} The Ricci tensor can be calculated from eq.(\ref{Ricciop}). Using eq.(\ref{NilsolitonRicci}) the derivation ${\bf D}$ can then be determined. 
\end{enumerate}
An important observation is that the eigenvalues of ${\bf D}$, up to a scalar multiplication, can be arranged into a tuple:
\beq
(k;d)=(k_1<k_2<...<k_r;d_1,d_2,...,d_r),
\eeq
where the $k_l$ are integers without common divisors and $d_l$ are their corresponding multiplicities. This tuple is called the eigenvalue type. Usually the Ricci nilsolitons are given in terms of its eigenvalue type as above due to the relation to the classification of Einstein solvmanifolds (see, e.g, \cite{Heber}). 

If $\mu$ is an $n$-dimensional nilpotent Lie algebra for which $\mu/\langle\mu,\mu\rangle^{1/2}$ corresponds to a fixed point of $F(\mu)$ as explained above, then 
\beq
{\bf D}_{\mu}=\frac{\langle\mu,\mu\rangle^{\frac 12}}{2}\left[n\left(k_1^2d_1+...+k_r^2d_r\right)-(k_1d_1+...+k_rd_r)^2\right]^{-\frac 12}\widetilde{\bf D},
\eeq
where $\widetilde{\bf D}$ is the derivation of $\mu$ with eigenvalues $k_i$ of multiplicities $d_i$. 

Let $\tilde\mu$ be the extension of $\mu$ by adding an Abelian factor: $\tilde\n=\n\oplus\mathbb{R}^m$; i.e., $\left.\tilde\mu\right|_{\n\times\n}=\mu$ and $[\mathbb{R}^m,\n]=0$. Then $F(\tilde\mu)=F(\mu)$ and the critical point has the eigenvalue type
\beq
\left(\alpha k_1<...<\frac{k_1^2d_1+...+k_r^2d_r}{d}<...<\alpha k_r;d_1<...<m<...<d_r\right),
\eeq
 where $d=\mathrm{mcd}(k_1d_1+...+k_rd_r,k_1^2d_1+...+k_r^2d_r)$ and $\alpha=\frac{k_1d_1+...+k_rd_r}{d}$. In the case that $\frac{k_1^2d_1+...+k_r^2d_r}{d}=\alpha k_i$ for some $i$, then the multiplicity is $m+d_i$. This result will be useful for us since adding a time-direction will add an additional one-dimensional Abelian factor to the nilpotent group. 

\section{From Ricci nilsolitons to black holes}
The reason for stressing the properties of the Ricci nilsolitons is because of the importance it has for constructing Einstein solvmanifolds. This relation between Ricci nilsolitons and Einstein solvmanifolds seems to have been noticed by Lauret \cite{L1,L2,L3,L4}. Moreover, when going from the homogeneous solvmanifold to the inhomogeneous black hole solutions, the isometry group of the Ricci nilmanifolds will survive the construction of the black hole spacetime. The Ricci flow will therefore have a particular role  for the black hole spacetimes\footnote{Interestingly, the role of the Ricci flow and black holes has been studied in a different context in \cite{Wiseman}.}.

\subsection{Einstein solvmanifolds and Ricci nilsolitons} 
Let us first state a theorem due to Lauret \cite{L1}:
\paragraph{Theorem [Lauret]}:
{\sl A homogeneous nilmanifold $(\n,\langle\cdot,\cdot\rangle)$ is a Ricci nilsoliton if and only if $(\n,\langle\cdot,\cdot\rangle)$ admits a metric solvable extension $(\s=\mf{a}\oplus\n,{\bf g})$ with $\mf{a}$ Abelian whose corresponding solvmanifold $(S,{\bf g})$ is Einstein.}
\paragraph{}
 This solvmanifold can be constructed as follows: Consider the following metric solvable extension of the nilpotent algebra $\n$ (with brackets $\mu$):
 \beq
\s=\mathfrak{a}\oplus\n,\quad [\s,\s]_{\s}=\n, \quad [\cdot,\cdot]_{\s}\big{|}_{\n\times\n}=[\cdot,\cdot]_{\n},
\eeq
and 
\beq
\langle\cdot,\cdot\rangle_{\s}\big{|}_{\n\times\n}=\langle\cdot,\cdot\rangle_{\n}, \quad \langle\mathfrak{a},\n\rangle_{\s}=0.
\eeq
Moreover, let $\mathfrak{a}=\mathbb{R}$, ${\bf D}\in \mathrm{Der}(\mu)$ and let ${\sf E}\in\mathfrak{a}$ such that $\langle{\sf E},{\sf E}\rangle_{\s}=1$. Then we can define the solvable Lie algebra by 
\beq
[{\sf E},{\sf e}_{i}]={\bf D}{\sf e}_i, \quad [{\sf e}_i,{\sf e}_j]=\mu({\sf e}_i,{\sf e}_j). 
\eeq
Let $(S,{\bf g})$ be the corresponding solvmanifold equipped the left-invariant metric. The Ricci tensor of $S$ can found to be (\cite{Kaplan})
\beq
{\bf g}({\bf R}_{\s}{\sf E},{\sf E})&=&-\Tr({\bf D}^2), \\
{\bf g}({\bf R}_{\s}{\sf E},{\sf e}_i)&=&0, \\
{\bf g}({\bf R}_{\s}{\sf e}_i,{\sf e}_j)&=&{\bf g}([-\Tr({\bf D}){\bf D}+{\bf R}_{\mu}]{\sf e}_i,{\sf e}_j).
\eeq
Hence, $(S,{\bf g})$ is Einstein if $\mu$ is a Ricci nilsoliton with ${\bf D}\in\textrm{Der}(\mu)$ given in eq.(\ref{NilsolitonRicci}):
\beq
{\bf R}_{\s}=c_{\mu}{\bf 1}, \quad c_{\mu}=-\Tr({\bf D}^2).
\eeq 
These Einstein solvmanifolds will correspond to the asymptotic metric for the black holes. In particular, this means that the black holes are not asymptotically AdS, but rather asymptotically a solvmanifold. 

The Einstein solvmanifolds has been in the centre for a long outstanding question regarding the classification of negatively curved homogeneous Einstein manifolds \cite{Besse,Heber}. All known examples of negatively curved homogeneous Einstein manifold are isometric to a Einstein solvmanifold, however, it is not proven that all necessarily are. Only in low dimensions some progress had been made (see \cite{Nikonorov} for dimension 5)\footnote{Moreover, recently a paper by Lauret appeared \cite{L07} proving that all Einstein solvmanifolds are necessary standard (see also \cite{LW}).}. In spite of the lack of a classification result, numerous examples of Einstein solvmanifolds exist. The most commonly known are the real, the complex and the quaternionic hyperbolic spaces $\H^n$, $\H_\C^n$, $\H_{\H}^n$,  and the Cayley hyperbolic plane $\H^2_{{\sf Cay}}$. These hyperbolic spaces can be further generalised to the so-called Damek-Ricci spaces \cite{GHG} which are solvable extensions of generalised Heisenberg spaces. 

\subsection{Black holes} 
We can now proceed to  constructing  Ricci nilsoliton black holes. More specifically, the constructed black hole spacetime, in spite of being  inhomogeneous, will possess an isometry group inherited from the Ricci nilsolitons. These black hole solutions were discussed in an earlier paper in a more general context \cite{Sig}. For the time being it is advantageous to keep the manifold Riemannian and assume that you can foliate the space using Ricci nilsoliton hypersurfaces. We introduce the extrinsic curvature ${\bf k}$ which is a bilinear and symmetric tensor living on the hypersurfaces. We define the extrinsic curvature operator ${\bf K}:~\n\mapsto\n$ by 
\[\langle{\bf K}{\sf e}_i,{\sf e}_j\rangle={\bf k}({\sf e}_i,{\sf e}_j).\]
Let us also introduce the Gaussian coordinate $y$ such that $\partial/\partial y$ is a unit normal vector to the nilmanifolds. Assume further that $\n$ contains an Abelian factor spanned by ${\sf e}_1$, say, so that $[{\sf e}_1,\n]=0$. This Abelian factor will eventually correspond to the time direction. For the Ricci nilsoliton this implies \beq
{\bf R}_{\mu}{\sf e}_1=0.
\label{Re=0}
\eeq  
Moreover, let ${\bf D}$ be the constant derivation given earlier. We can decompose ${\bf D}$ and ${\bf K}$ as 
\beq
{\bf D}=\begin{bmatrix}
D_{11}& 0 \\
0 & \widetilde{\bf D}
\end{bmatrix}, \quad 
{\bf K}=\begin{bmatrix}
K_{11}& 0 \\
0 & \widetilde{\bf K}
\end{bmatrix}.
\eeq
Here, $D_{11}=\Tr({\bf D}^2)/\Tr({\bf D})$, which follows from eq.(\ref{Re=0}). 
By assuming $\widetilde{{\bf K}}=\Lambda(y)\widetilde{{\bf D}}$ where $\Lambda(y)$ is some function of $y$, implies that ${\bf K}$ is also a derivation of $\n$. Since the derivations are the generators of the automorphism group, this choice of ${\bf K}$ implies that the geometry of the hypersurfaces is preserved as you go along the Gaussian coordinate $y$. So $\dot{\bf R}_{\mu}=0$, where dot denotes (Lie) derivative with respect to $y$, and hence, ${\bf R}_{\mu}=-\Tr({\bf D}^2){\bf 1}+\Tr({\bf D}){\bf D}$ where the derivation ${\bf D}$ can be considered to be a constant. 

The Gauss' equations now reduce to 
\beq
\dot{\bf K}+\Tr({\bf K}){\bf K}-{\bf R}_{\mu}+\lambda{\bf 1}&=&0, \\
\Tr({\bf K}^2)-[\Tr({\bf K})]^2+\Tr({\bf R}_\mu)-(n-1)\lambda&=& 0.
\eeq
We note first that the solution given by 
\beq
{\bf K}={\bf D}
\eeq 
is the Einstein solvmanifold given above, with $\lambda=-\Tr({\bf D}^2)$.

Another set of solutions can be found by 
\beq
{\bf K}=\coth[D(y-y_0)]{\bf D}+\frac{{\mbold\sigma}}{\sinh[D(y-y_0)]},
\label{eq:Ksol}\eeq
where we have set $D=\Tr({\bf D})$ and
\beq
{\mbold\sigma}=\begin{bmatrix}
(D-D_{11})& 0 \\
0 & -\widetilde{{\bf D}}
\end{bmatrix}.
\eeq
We note that ${\mbold\sigma}$ is trace-free and orthogonal to ${\bf D}$; i.e.,
\beq
\Tr({\mbold\sigma})=0, \quad \Tr({\mbold\sigma}{\bf D})=0.
\eeq
The ${\bf R}_{\mu}$ and $\lambda$ are constants and given as above.

The solutions given above are the Euclidean versions of Ricci nilsoliton black holes given in terms of the nilsoliton foliation of the solutions. It is useful to write down the metric for this solution in the standard form. This can be accomplished by introducing the coordinate $w$ by 
\beq
y-y_0=\frac{2}{D}\mathrm{artanh}\sqrt{1-M\exp(-Dw)}.
\eeq
By diagonalising ${\bf D}=\mathrm{diag}(q_1,q_2,...,q_n)$, we can write 
\beq
\mathrm{d} s^2=\frac{\mathrm{d} w^2}{1-Me^{-Dw}}+(1-Me^{-Dw})e^{2q_1w}(\mathrm{d} x^1)^2+\sum_{i=2}^ne^{2q_iw}\left({\mbold\omega}^i\right)^2,
\eeq
where $\{\mathrm{d} x^1,{\mbold\omega}^i\}$ is an appropriate set of left-invariant vectors on $\n$. These obey $\mathrm{d} {\mbold\omega}^k=-(1/2)C^k_{ij}{\mbold\omega}^i\wedge{\mbold\omega}^j$ where $C^k_{ij}$ are constants and are the structure constants of $\n$. 

A Lorentzian  solution can now be found by Wick-rotating the coordinate $x^1$; i.e., by setting $t=ix^1$, we get
\beq
\mathrm{d} s^2=-(1-Me^{-Dw})e^{2q_1w}\mathrm{d} t^2+\frac{\mathrm{d} w^2}{1-Me^{-Dw}}+\sum_{i=2}^ne^{2q_iw}\left({\mbold\omega}^i\right)^2.
\eeq
A more standard form can be accomplished by defining a new variable $r$ by $w=(1/q_1)\ln (q_1r)$ for which we get 
\beq
\mathrm{d} s^2=-f(r)\mathrm{d} t^2+\frac{\mathrm{d} r^2}{f(r)}+h_{AB}(r){\mbold\omega}^A{\mbold\omega}^B, \quad f(r)=q_1^2r^2-{M}{(q_1r)^{-\frac{D-2q_1}{q_1}}}.
\eeq
We therefore see that these nilsoliton black holes are generalisation of the standard toroidal AdS black holes. This is also clear from the fact that the toroidal black holes have flat horizon geometry; so in a sense, the toroidal AdS black hole is the trivial case where the nilpotent group is Abelian.

We usually assume that black holes have compact horizons. In order for the horizon to be compact, one must require that the nilmanifolds allow for a compact quotient; i.e., there exists a lattice $\Gamma\subset N_{\mu}$ such that $N_{\mu}/\Gamma$ is a compact manifold. For nilmanifolds the existence of such a lattice can be determined using the Lie algebra $\mu$  \cite{Eberlein}: 
\paragraph{Theorem:}{\sl A nilmanifold can be compactified if and only if there exists a frame such that $[{\sf e}_i,{\sf e}_j]=C^k_{ij}{\sf e}_k$ where $C^{k}_{ij}$ are all rational constants. }
\paragraph{}
By inspection of the nilpotent Lie algebras of dimension $\leq 6$ we get an immediate consequence of this theorem: 
\paragraph{Corollary:} {\sl All nilmanifolds of dimension $\leq 6$ allow for a compact quotient.} 
\paragraph{}
There are only a finite number of nilpotent Lie algebras of dimension $\leq 6$ (50 including the Abelian ones), all of which allow for a Ricci nilsoliton metric \cite{Will} (a copy of Will's list is given in the Appendix). 
 
Among the 7-dimensional Lie algebras, there exists a curve of non-isometric Lie algebras. These allow for a nilsoliton metric of type (see \cite{L3})
\[ (1<2<3<4<5<6<7; 1,...,1).\]
The curve of nilsoliton Lie algebras can be given by 
\beq
&&\mu^3_{12}=\mu^7_{34}=(1-t)^{\frac 12}, \quad \mu^5_{14}=\mu^7_{23}=t^{\frac 12}\nonumber \\
&& \mu^4_{13}=\mu^6_{15}=\mu^7_{16}=\mu^5_{23}=\mu^6_{24}=1.
\eeq
This algebra can be shown to be isomorphic to the algebra denoted 1,2,3,4,5,7$_I$:$t$ in Seeley's list of 7-dimensional nilpotent Lie algebras \cite{Seeley}. Hence, the above algebra is isomorphic to the Lie algebra given by 
\beq
&&\tilde{\mu}^7_{34}=(1-t), \quad \tilde{\mu}^7_{23}=t\nonumber \\
&& \tilde{\mu}^3_{12}=\tilde{\mu}^5_{14}=\tilde{\mu}^4_{13}=\tilde{\mu}^6_{15}=\tilde{\mu}^7_{16}=\tilde{\mu}^5_{23}=\tilde{\mu}^6_{24}=1.
\eeq
Thus, by virtue of the above theorem, if $t$ is rational then the corresponding nilsoliton metric allows for a compact quotient. This implies that there exists an infinite number of model nilmanifolds which allows for a compact quotient. 

Therefore, if we classify the black hole solutions in terms of the model geometries, we have that for every nilpotent group of dimension $\leq 6$, there exists a corresponding black hole solution in dimension $\leq 8$. Moreover, for any dimension $>8$, there is an infinite number of locally distinct black hole solutions with a nilsoliton metric as a horizon.  

Note that there may be many different lattices $\Gamma$ for a given nilmanifold (for example, there is an infinite number of possible non-homeomorphic quotients of $N_{3,1}$). Also, we have not addressed the issue of moduli space of non-isometric quotients. 

\subsection{Making the Euclidean solution regular}
To make the Euclidean solution regular we must ensure that the solution behaves regularly at the horizon. We can use the Gaussian coordinate $y$ and approximate the solution close to $y=y_0$. This yields 
\beq
\mathrm{d} s^2\approx \mathrm{d} y^2+(y-y_0)^2\left(\frac{D}{2}M^{\frac{q_1}{D}}\mathrm{d} x^1\right)^2+\sum_{i=2}^nM^{\frac{2q_i}{D}}({\mbold\omega^i})^2.
\eeq
Hence, if we identify $x^1$ under the map 
\[ x^1\longmapsto x^1+\frac{4\pi}{DM^{\frac{q_1}D}},\]
the solution closes of regularly and the Euclidean solution is everywhere regular.

If we write this identification as $x^1\mapsto x^1+\beta$ then $\beta$ is usually interpreted as the inverse temperature of the black hole; i.e., $\beta=1/T$. This implies that $T\propto M^{q_1/D}$ and so the temperature increases as the mass increases. 
\subsection{Generalisations} 
Let us recapitulate what assumptions were made in order for eq.(\ref{eq:Ksol}) to be a solution: 
\begin{enumerate}
\item{} ${\bf K}$ is a derivation of $\n$.
\item{} $\Tr(\mbold\sigma)=\Tr(\mbold\sigma{\bf D})=0$.
\item{} $\Tr(\mbold\sigma^2)=[\Tr({\bf D})]^2-\Tr({\bf D}^2)$.
\end{enumerate}
We can therefore generalise the above solution as long as these criteria are satisfied.

So, for example, consider an $m$-dimensional Abelian factor of $\n$ such that $[\mathbb{R}^m,\n]=0$. Furthermore, assume that the Abelian factor is spanned by ${\sf e}_1,...,{\sf e}_m$. Then $\langle{\bf R}_{\mu}{\sf e}_i,{\sf e}_i\rangle=0$ for $i=1,...,m$ and the derivation ${\bf D}$ can be decomposed as
\beq
{\bf D}=\begin{bmatrix}
D_{11}{\bf 1}_{m\times m}& 0 \\
0 & \widetilde{\bf D}
\end{bmatrix}.
\eeq
Then eq.(\ref{eq:Ksol}) is a solution as long as we choose ${\mbold\sigma}$ the following way:
\beq
{\mbold\sigma}=\begin{bmatrix}
\frac 1m(D-mD_{11}){\bf 1}_{m\times m}+{\bf A}& 0 \\
0 & -\widetilde{\bf D}
\end{bmatrix},\eeq
where the $m\times m$ matrix ${\bf A}$ obeys
\beq
\Tr({\bf A})=0, \quad \Tr({\bf A}^2)=\frac{m-1}{m}D^2.
\eeq
Since the solutions are only defined locally, it is not clear what the interpretation of these solutions are or whether they can be made regular by an appropriate identification. 

\section{Properties of Ricci nilsoliton Black Holes} 
Let us consider the Lorentzian Ricci nilsoliton black hole metric:
\beq
\mathrm{d} s^2=-(1-Me^{-Dw})e^{2q_1w}\mathrm{d} t^2+\frac{\mathrm{d} w^2}{1-Me^{-Dw}}+\sum_{i=2}^ne^{2q_iw}\left({\mbold\omega}^i\right)^2.
\eeq
First we note that $w=\infty$ corresponds to an infinite value of the Gaussian coordinate $y$; thus spatial infinity is infinitely far away. Moreover, the horizon is located at $y=y_0$ which corresponds to $w=(1/D)\ln M$. 

\subsection{Geodesics}
For outbound null-geodesics travelling in the $w$-direction we get
\beq
\frac{\mathrm{d} t}{\mathrm{d} w}=\frac{e^{-q_1w}}{1-Me^{-Dw}}.
\eeq
So by integration
\beq
t-t_0=\int_{w_0}^w\frac{e^{-q_1w}\mathrm{d} w}{1-Me^{-Dw}}\leq \int_{w_0}^w{e^{-q_1w}}\mathrm{d} w\leq \frac{e^{-q_1w_0}}{q_1},
\eeq
and hence, $t-t_0$ is bounded. This implies that light-rays reach spatial infinity within finite coordinate time. In this way light-signals can leak through spatial infinity. This is analogous to the  Anti-de Sitter spacetime. 

Consider now timelike geodesics and let $p_t$ be the canonical momentum of $t$: 
\beq
p_t=\frac{\partial L}{\partial \dot{t}}=\mathrm{constant}.
\eeq
Then, by using the identity $g_{\mu\nu}\dot{x}^{\mu}\dot{x}^{\nu}=-1$, we get 
\beq
\dot{w}^2=p_t^2e^{-2q_1 w}-(1-Me^{-Dw})\left(1+h_{ab}\dot{x}^a\dot{x}^b\right)\leq p_t^2e^{-2q_1 w}-(1-Me^{-Dw}).
\eeq
Hence, for any timelike geodesic, there exists a hypersurface given by $w=w_{\mathrm{max}}$ for which the geodesic can never pass; i.e., $w(\tau)\leq w_{\mathrm{max}}$. The outward going geodesics ultimately stop and start to go inwards. 
Any timelike geodesic will ultimately cross the horizon at some time in the future. 

\subsection{The mass of the black hole}
Analogous to the Ashtekar-Magnon-Das conformal mass, we can define the 'mass' of the black hole as follows (see, e.g., \cite{DK}): 
\beq
\widetilde{M}=-\lim_{w\rightarrow \infty}\oint m^{\rho}m^{\sigma}\left(R_{\mu\rho\sigma\nu}-\overline{R}_{\mu\rho\sigma\nu}\right)n^{\mu}\xi^{\nu}\mathrm{d} {\bf S}. 
\eeq
Here, the $\mathrm{d} {\bf S}$ is the volume element of the surfaces defined by the Ricci nilsolitons; $m^{\mu}$ and $n^{\mu}$ are orthonormal vectors orthogonal to $S$: 
\[ g^{\mu\nu}n_{\mu}n_{\nu}=-1, \quad g^{\mu\nu}m_{\mu}m_{\nu}=1, \quad g^{\mu\nu}m_{\mu}n_{\nu}=0;\]
$\xi^{\mu}$ is the timelike Killing vector field $\partial_t$; $R_{\mu\rho\sigma\nu}$ is the Riemann tensor of the black hole metric; and $\overline{R}_{\mu\rho\sigma\nu}$ is the Riemann tensor of 'background' Einstein solvmanifold (this will be justified later). 

For the Ricci nilsoliton black holes, we get
\beq
\widetilde{M}=\frac 12M(D-2q_1)(D-q_1)\mathrm{Vol}(N_{\mu}/\Gamma),
\eeq
where $\mathrm{Vol}(N_{\mu}/\Gamma)$ is the volume of the compact hypersurfaces at $w=0$. This shows that the parameter $M$, up to a constant, can indeed be interpreted as a 'mass' relative to the background solvmanifold.

\subsection{The asymptotic geometry} 
Consider the (Euclidean) spacetime close to spatial infinity, $y=\infty$. The extrinsic curvature can be approximated by 
\beq
{\bf K}={\bf D}+2\mbold{\sigma}e^{-D(y-y_0)}+\mathcal{O}(e^{-2D(y-y_0)}).
\eeq
Hence, asymptotically, the spacetime approaches the corresponding solvmanifold as claimed. The solvmanifold spacetime can therefore be considered as the background spacetime in which there is a black hole. Sufficiently far away from the black hole, the spacetime can be approximated as a solvmanifold.   
These solutions are therefore black hole solutions with a negative cosmological constant, which are not (locally) asymptotically Anti-de Sitter. 

In the AdS case, the isometry group acts on the conformal boundary as conformal transformations. This is directly related to the fact that for real hyperbolic space, which is the Euclidean version of AdS space, we have the relation \cite{BP}
\beq
\mathrm{Isom}(\H^n)=\mathrm{Conf}(\partial \H^n),
\label{Isom=Conf}\eeq
where $\partial \H^n$ is the conformal boundary of $\H^n$. The conformal boundary of  $\H^n$ can be identified as the one-point compactification of flat space, $\E^{n-1}$; hence, we can write $\mathrm{Isom}(\H^n)=\mathrm{Conf}(\E^{n-1})$. This relation lies as a foundation of many works on AdS space. 

For the nilsoliton black holes, we believe there is an analogous (but not identical) relation for the asymptotic geometry. Firstly, it is easy to see that the isometries of the horospheres are preserved as $y\rightarrow \infty$. Secondly, the derivation, ${\bf D}$, generates a one-parameter group of automorphisms acting on the asymptotic nilsolitonic geometry. This one-parameter group of automorphisms can be viewed as a \emph{dilaton}, $\phi_t$, acting on the nilsolitons. This dilaton and the isometries of the nilsolitons act transitively on the 'background' solvmanifold. In addition to these symmetries, the background solvmanifold may possess some additional isometries. In the case of $\H^n$, these additional symmetries act as 'inversions' on the conformal boundary $\E^{n-1}$ \cite{BP}. 

\subsubsection{Complex hyperbolic space}
For a general solvmanifold, it is not known whether there exists a similar identity as (\ref{Isom=Conf}). However, let us consider the complex hyperbolic space, $\H_{\C}^n$, where a related relation is known to exist. Let us also, for simplicity, restrict to 2 complex dimensions  even though the following can easily be generalised to  $\H_{\C}^n$ for any $n$.  

For $\H_\C^2=SU(1,2)/U(2)$ the group $SU(1,2)$ acts as isometries with $U(2)$ as an isotropy group. For $\H_\C^2$ the boundary can be considered to be the one-point compactification of the Heisenberg group $\mathrm{Heis}_3$. Let us introduce the coordinates $(x,y,z)$ on $\mathrm{Heis}_3$ such that the one-forms
\beq
{\mbold\omega}=\mathrm{d} x+2(y \mathrm{d} z-z \mathrm{d} y), \quad \mathrm{d} y, \quad \mathrm{d} z,
\eeq
are the left-invariant one-forms on $\mathrm{Heis}_3$. \footnote{The one-form ${\mbold\omega}$ defines a \emph{contact structure} on $\mathrm{Heis}_3$, i.e., ${\mbold\omega}\wedge(\mathrm{d} {\mbold\omega})^m$, $m\in\mathbb{N}$ is the volume form. A transformation $f:~M\mapsto M$ is a \emph{contact transformation} if
\beq
f^{*}{\mbold\omega}=\lambda{\mbold\omega},
\eeq
for a scalar function $\lambda$. The group action of $SU(2,1)$ will act on $\mathrm{Heis}_3$ as contact transformations. However, the contact structure does not generalise to all the nilpotent black hole spacetimes.} By considering the dual vector to ${\mbold\omega}$, ${\sf e}_x$, we see that ${\sf e}_x\in\mf{g}_D^{(1)}$ where $\mf{g}_D^{(i)}$ is the  derived series of the Heisenberg algebra. Moreover, $\mf{g}_D^{(2)}=0$ so  $\mf{g}_D^{(1)}$ is abelian. 

Consider therefore a general nilpotent Lie algebra, $\n$, and let $\mf{g}_D^{(i)}$ be its  derived series. Moreover, let $k$ be the largest number such that $\mf{g}_D^{(k)}\neq 0$ and $\mf{g}_D^{(k+1)}=0$. This implies that $\mf{g}_D^{(k)}$ is an abelian ideal in $\n$ and is in the center of $\n$. By considering a left-invariant ${\sf e}\in\mf{g}_D^{(k)}$, and using eq.(\ref{Ricciop}), we note that,  
\beq 
\langle{\bf R}_\mu{\sf e},{\sf e}\rangle=\frac 14\sum_{kl}\langle\mu({\sf e}_k,{\sf e}_l),{\sf e}\rangle^2\geq 0,
\eeq
where $=0$ if and only if $\n$ is Abelian. This implies, using eq.(\ref{NilsolitonRicci}), that the biggest eigenvalues for ${\bf D}$ will correspond to the  the ideal $\mf{g}_D^{(k)}$. Hence, the ideal $\mf{g}_D^{(k)}$ will dominate the asymptotic geometry of the solvmanifold. This ideal therefore plays an important role for the asymptotic geometry; more specifically, assume that $q_n$ is the largest eigenvalue of ${\bf D}$, then the conformal transformation 
\beq
\mathrm{d} s^2\mapsto e^{-2q_nw}\mathrm{d} s^2, 
\eeq
renders the limit ${\bf g}_{\partial}\equiv \lim_{w\rightarrow \infty}e^{-2q_nw}\mathrm{d} s^2$ well defined. We can therefore study the symmetry group of the solvmanifold acting on the symmetric two-tensor ${\bf g}_{\partial}$ which lives on the boundary. 

Another possible generalisation is the following observation for complex hyperbolic spaces (see, e.g., \cite{KR,AM}). Define the following 'gauge' on $\mathrm{Heis}_3$:
\beq
\| g\| =\left[x^2+(y^2+z^2)^2\right]^{\frac 14}, \quad \text{where}~~g=(x,y,z)\in\mathrm{Heis}_3.
\label{gauge}\eeq
We define the left-invariant distance between $g$ and $g'$  by 
\beq
d_H(g,g')=\|g^{-1}g'\|.
\eeq 
We note that this metric is \emph{not Riemannian}. On the other hand, we do note that $SU(2,1)$ acts conformally with respect to this metric; in fact, 
\beq
\mathrm{Isom}(\H^2_\C)=\mathrm{Conf}_{d_H}(\mathrm{Heis}_3).
\eeq
By introducing eq.(\ref{gauge}) we can thus manifestly generalise eq.(\ref{Isom=Conf}) to complex hyperbolic spaces. The key observation is that the gauge preserves the symmetries of Heis$_3$ and that the dilaton acts homogeneously on the gauge.  

We can speculate whether either of these paths can be followed to generalise eq.(\ref{Isom=Conf}) to the black hole spacetimes considered here. More work is clearly needed here. 

\section{Discussion} 
Here we have discussed how we can construct Ricci nilsoliton black holes from nilpotent groups. The corresponding black hole spacetimes are solutions to Einstein's equations with a negative cosmological constant. We have given conditions for when such solutions exists and, in particular, we have shown that any nilpotent group of dimension $\leq 6$ has a corresponding Ricci nilsoliton black hole in dimension $\leq 8$. In dimensions higher than 8, there are, in each dimension, an infinite number of locally distinct Ricci nilsoliton black hole spacetimes. 

The lowest dimension where there exists a non-trivial nilpotent group is 3. In this regard, Cadeau and Woolgar \cite{CW} seem to be the first to construct the corresponding black hole in dimension 5. However, apart from this solution, the nilpotent black holes seem to have gone unnoticed in the literature (they are also pointed out by the author in \cite{Sig} but in a more general context). 

The negatively curved spaces seem to have an incredible rich structure, some of which are displayed in this work. This rich structure makes the negatively curved spaces difficult to study in general which is probably the main reason for the lack of understanding  of such spaces. However, on the same token, the wealth of different phenomena these spaces possess is also what makes them so interesting\footnote{Another application of solvmanifolds can be seen in \cite{solwaves}.}. It is clear that we only have unveiled  the tip of the iceberg and that many more treasures remain to be discovered.

\section*{Acknowledgments}
 This work was supported by an AARMS Postdoctoral Fellowship.  

\appendix
\section{Ricci nilsolitons of low dimension}
\label{App}
The following list contain all nilpotent Lie algebras with their critical points. The eigenvalue type is also included. These tables are taken from \cite{L3,Will}. 

The notation used is best illustrated with an example. The tuple 
\[ (0,0,\sqrt{3}[12],\sqrt{3}[13],\sqrt{2}[14]+\sqrt{2}[23])\] 
represents the Lie algebra
\beq
&& [{\sf e}_1,{\sf e}_2]=\sqrt{3}{\sf e}_3, \quad [{\sf e}_1,{\sf e}_3]=\sqrt{3}{\sf e}_4, \nonumber    \\ 
&& [{\sf e}_1,{\sf e}_4]=\sqrt{2}{\sf e}_5, \quad [{\sf e}_2,{\sf e}_3]=\sqrt{2}{\sf e}_5. \nonumber 
\eeq
\subsection{Dimension 3}
\begin{tabular}{|l|c|c|c|}
\hline
& Critical point & Eigenvalue type & comments\\
\hline
\hline
$N_{3,1}$ & $(0,0,[12])$ & $(1<2;2,1)$ & $\mathrm{Heis}_3$  \\
\hline
\end{tabular}
\subsection{Dimension 4}
\begin{tabular}{|l|c|c|c|}
\hline
& Critical point & Eigenvalue type & comments\\
\hline
\hline
$N_{4,1}$ & $(0,0,[12],[13])$ & $(1<2<3<4;1,1,1,1)$ &   \\
\hline
$N_{4,2}$ & $(0,0,[12],0)$ & $(2<3<4;2,1,1)$ & $\mathrm{Heis}_3\oplus\mathbb{R}$  \\
\hline
\end{tabular}
\subsection{Dimension 5}
\begin{tabular}{|l|c|c|c|}
\hline
& Critical point & Eigenvalue type & comments\\
\hline
\hline
$N_{5,1}$ & $(0,0,3[12],4[13],3[14])$ & $(2<9<11<13<15;1,...,1)$ &   \\
\hline
$N_{5,2}$ & $(0,0,\sqrt{3}[12],\sqrt{3}[13],\sqrt{2}[14]+\sqrt{2}[23])$ & $(1<2<3<4<5;1,...,1)$ &   \\
\hline
$N_{5,3}$ & $(0,0,0,[12],\sqrt{2}[14]+\sqrt{2}[23])$ & $(3<4<6<7<10;1,...,1)$ &   \\
\hline
$N_{5,4}$ & $(0,0,0,0,[12]+[34])$ & $(1<2;4,1)$ &  $\mathrm{Heis}_5$ \\
\hline
$N_{5,5}$ & $(0,0,4[12],3[13],3[23])$ & $(1<2<3;2,1,2)$ &   \\
\hline
$N_{5,6}$ & $(0,0,0,[12],[13])$ & $(2<3<5;1,2,2)$ &   \\
\hline
$N_{5,7}$ & $(0,0,0,0,[12])$ & $(2<3<4;2,2,1)$ &  $\mathrm{Heis}_3\oplus\mathbb{R}^2$ \\
\hline
$N_{5,8}$ & $(0,0,0,[12],[14])$ & $(1<2<3<4;1,1,2,1)$ &  $N_{4,1}\oplus\mathbb{R}$ \\
\hline
\end{tabular}
\subsection{Dimension 6} 
\subsubsection{5 and 4 step:}
\small
\begin{tabular}{|l|c|c|}
\hline
& Critical point & Eigenvalue type \\
\hline
\hline
$N_{6,1}$ & $\begin{matrix}(0,0,\sqrt{13}[12],4[13],\\ 
			\sqrt{12}[14]+2[23],\sqrt{12}[34]+\sqrt{13}[52])\end{matrix}$ & $(1<2<3<4<5<7;1,...,1)$   \\
\hline
$N_{6,2}$ & $(0,0,[12],\sqrt{\frac 43}[13],[14],[34]+[52])$ & $(1<3<4<5<6<9;1,...,1)$   \\
\hline
$N_{6,3}$ & $(0,0,2[12],\sqrt{6}[13],\sqrt{6}[14],2[15])$ & $(1<9<10<11<12<13;1,...,1)$   \\
\hline
$N_{6,4}$ & $\begin{matrix}(0,0,\sqrt{22}[12],6[13],\\
			\sqrt{22}[14]+\sqrt{30}[23],5[24]+\sqrt{30}[15])\end{matrix}$ & $(1<2<3<4<5<6;1,...,1)$   \\
\hline
$N_{6,5}$ & $\begin{matrix}(0,0,\sqrt{7}[12],\sqrt{\frac{15}{2}}[13],\\
3[14],\sqrt{\frac{15}{2}}[23]+2[15])\end{matrix}$ & $(1<3<4<5<6<7;1,...,1)$   \\
\hline
$N_{6,6}$ & $(0,0,[12],[13],[23],[14])$ & $(1<2<3<4<5;1,1,1,1,2)$   \\
\hline
$N_{6,7}$ & $\begin{matrix}(0,0,2[12],\sqrt{5}[13],\\ 
		\sqrt{5}[23],2[14]-2[25])\end{matrix}$ & $(1<2<3<4;2,1,2,1)$  \\
\hline
$N_{6,8}$ & $\begin{matrix} (0,0,2[12],\sqrt{5}[13],\\
		\sqrt{5}[23],2[14]+2[25])\end{matrix}$ & $(1<2<3<4;2,1,2,1)$ \\
\hline
$N_{6,9}$ & $\begin{matrix}(0,0,0,\sqrt{\frac 54}[12],\\
		[14]-[23],\sqrt{\frac 54}[15]+[34])\end{matrix} $ & $(6<11<12<17<23<29;1,...,1)$  \\
\hline
$N_{6,10}$ & $(0,0,0,[12],\sqrt{\frac 53}[14],[15]+[23])$ & $(4<9<12<13<17<21;1,...,1)$ \\
\hline
$N_{6,11}$ & $\begin{matrix} (0,0,-\sqrt{\frac{35}{136}}[12],\sqrt{\frac{21}{34}}[12],\\ 
\sqrt{\frac{25}{68}}[14]-\sqrt{\frac{15}{17}}[13],\sqrt{\frac {3}{4}}[15]+\sqrt{\frac 78}[24])\end{matrix}$ & $(1<2<3<4<5;1,1,2,1,1)$ \\
\hline
$N_{6,12}$ & $(0,0,0,\sqrt{3}[12],\sqrt{3}[14],\sqrt{2}[15]+\sqrt{2}[24]) $ & $(3<6<9<11<12;1,...,1)$  \\
\hline
$N_{6,13}$ & $(0,0,0,\sqrt{3}[12],2[14],\sqrt{3}[15])$ & $(2<9<11<12<13<15;1,...,1)$ \\
\hline
\end{tabular}
\subsubsection{3 and 2 step:}
\small
\begin{tabular}{|l|c|c|}
\hline
& Critical point & Eigenvalue type \\
\hline
\hline
$N_{6,14}$ & $(0,0,0,\sqrt{3}[12],\sqrt{2}[13],\sqrt{2}[14]+\sqrt{3}[35]) $ & $(2<3<4<5<6<8;1,...,1)$  \\
\hline
$N_{6,15}$ & $(0,0,0,[12],[23],[14]+[35])$ & $(1<2<3;3,2,1)$ \\
\hline
$N_{6,16}$ & $(0,0,0,[12],[23],[14]-[35])$ & $(1<2<3;3,2,1)$ \\
\hline
$N_{6,17}$ & $(0,0,0,2[12],\sqrt{3}[14],\sqrt{3}[24])$ & $(5<10<12<15;2,1,1,2)$ \\
\hline
$N_{6,18}$ & $\begin{matrix} (0,0,0,\sqrt{2}[12],\sqrt{\frac{1}{2}}[13]+\sqrt{\frac 32}[42],\\ 
\sqrt{\frac 32}[14]+\sqrt{\frac 12}[23])\end{matrix}$ & $(1<2<3;2,2,2)$ \\
\hline
$N_{6,19}$ & $(0,0,0,2[12],\sqrt{3}[14],[13]+\sqrt{3}[42])$ & $(5<6<11<12<16<17;1,...,1)$ \\
\hline
$N_{6,20}$ & $(0,0,-[12],\sqrt{3}[12],2[14],[24]-\sqrt{3}[23])$ & $(1<2<3;2,2,2)$ \\
\hline
$N_{6,21}$ & $(0,0,0,\sqrt{2}[12],[13],\sqrt{2}[14]+[23])$ & $(3<5<6<8<9;1,...,1)$ \\
\hline
$N_{6,22}$ & $(0,0,0,\sqrt{\frac 34}[12],\sqrt{\frac 34}[13],[24])$ & $(5<6<9<11<15<16;1,...,1)$ \\
\hline
$N_{6,23}$ & $(0,0,0,\sqrt{2}[12],[13],\sqrt{2}[14])$ & $(2<5<6<7<8<9;1,...,1)$ \\
\hline
$N_{6,24}$ & $(0,0,0,[12],[13],[23])$ & $(1<2;3,3)$ \\
\hline
$N_{6,25}$ & $(0,0,0,0,{2}[12],\sqrt{3}[15]+\sqrt{3}[34])$ & $(5<8<9<13<18;1,1,2,1,1)$ \\
\hline
$N_{6,26}$ & $(0,0,0,0,[12],[15])$ & $(1<2<3<4;1,1,3,1)$ \\
\hline
$N_{6,27}$ & $(0,0,0,0,\sqrt{2}[12],[14]+\sqrt{2}[25])$ & $(3<4<6<7<10;1,1,1,2,1)$ \\
\hline
$N_{6,28}$ & $(0,0,0,0.[13]+[42],[14]+[23])$ & $(1<2;4,2)$ \\
\hline
$N_{6,29}$ & $(0,0,0,0,[12],[14]+[23])$ & $(13<4<6<7;2,2,1,1)$ \\
\hline
$N_{6,30}$ & $(0,0,0,0,[12],[34])$ & $(1<2;4,2)$ \\
\hline
$N_{6,31}$ & $(0,0,0,0,[12],[13])$ & $(2<3<4<5;1,2,1,2)$ \\
\hline
$N_{6,32}$ & $(0,0,0,0,0,[12]+[34])$ & $(3<4<6;4,1,1)$ \\
\hline
$N_{6,33}$ & $(0,0,0,0,0,[12])$ & $(2<3<4;2,3,1)$ \\
\hline
\end{tabular}

\normalsize

We note the following Lie algebras contain an Abelian factor: 
\beq N_{6,12}=N_{5,2}\oplus\mathbb{R},&&  N_{6,13}=N_{5,1}\oplus\mathbb{R},\nonumber \\ 
N_{6,17}=N_{5,5}\oplus\mathbb{R}, && N_{6,26}=N_{4,1}\oplus\mathbb{R}^2, \nonumber \\
N_{6,27}=N_{5,3}\oplus\mathbb{R}, && N_{6,31}=N_{5,6}\oplus\mathbb{R}, \nonumber \\
N_{6,32}=N_{5,4}\oplus\mathbb{R}, && N_{6,33}=N_{3,1}\oplus\mathbb{R}^3. 
\eeq
Also worth noting are  the following isomorphisms: 
\beq
N_{6,28}=\mathrm{Heis}_{3,\mathbb{C}},\quad N_{6,30}=\mathrm{Heis}_3\oplus\mathrm{Heis}_3. 
\eeq

As an example, consider $N_{6,25}$. The components of the Ricci tensor can be found from eq.(\ref{Ricciop}):
\beq
{\bf R}_\mu=\mathrm{diag}\left(-\frac 72,-2,-\frac 32,-\frac 32,\frac 12,3\right),
\eeq
 which gives 
\beq
R(\mu)=-5, \quad F(\mu)=30.
\eeq
Since, ${\bf R}_{\mu}=c_{\mu}{\bf 1}+\Tr({\bf D}){\bf D}$, and $c_{\mu}=F(\mu)/R(\mu)$, we have
\beq
{\bf R}_{\mu}=-6\cdot {\bf 1}+\frac 12\mathrm{diag}(5,8,9,9,13,18).
\eeq
The eigenvalue value type is therefore given in the above table (which corrects a typo in \cite{Will}).

\end{document}